# Browser Extension for Fake URL Detection

Dr. Latesh G. Malik[1], Rohini Shambharkar[2], Shivam Morey [3], Shubhlak Kanpate[4] and Vedika Raut[5]
Department of Computer Science and Engineering
Government College of Engineering
Nagpur, India
Email: {[1]lgmalik,[2]rhshambharkar,[3]ssmorey, [4]srkanpate,[5]vsraut}@gcoen.ac.in

*Abstract*—In recent years, Cyber attacks have increased in number, and with them, the intensity of the attacks and their potential to damage the user have also increased significantly. In an ever-advancing world, users find it difficult to keep up with the latest developments in technology, which can leave them vulnerable to attacks. To avoid such situations we need tools to deter such attacks, for this machine learning models are among the best options. This paper presents a Browser Extension that uses machine learning models to enhance online security by integrating three crucial functionalities: Malicious URL detection, Spam Email detection and Network logs analysis. The proposed solution uses LGBM classifier for classification of Phishing websites, the model has been trained on a dataset with 87 features, this model achieved an accuracy of 96.5% with a precision of 96.8% and F1 score of 96.49%. The Model for Spam email detection uses Multinomial NB algorithm which has been trained on a dataset with over 5500 messages, this model achieved an accuracy of 97.09% with a precision of 100%. The results demonstrate the effectiveness of using machine learning models for cyber security.

## I. INTRODUCTION

The swift advancement of phishing assaults, spam emails, and network vulnerabilities has made it difficult for conventional detection and defence techniques to stay up to date. Phishing attacks have evolved to become more complex, frequently fooling even tech-savvy people and resulting in serious financial losses as well as the compromise of private information.Recently, hackers started doing their jobs very professionally, the recent phishing activity trends report [1] showed that 78% of all phishing websites use SSL protection that was exclusively used by authentic websites. Wandera stated in its 2020 Mobile Threat Landscape Report [2] that a new phishing website launches every 20 seconds.

Although spam emails remain the main means by which malware and phishing assaults are disseminated, attackers are constantly modifying their tactics to get around spam filters, which lowers productivity and increases the likelihood of successful attacks. Furthermore, network weaknesses allow infrastructure to be exploited, and manually analyzing network data is error-prone and labor-intensive.

Users are left vulnerable to ongoing threats in the event that there is no cohesive, flexible solution to handle these interconnected cybersecurity issues. Present methods frequently fall short of creating a unified, easily navigable tool that combines spam filtering, continuous network log analysis, and real-time bogus URL identification. Because of this weakness in the cybersecurity ecosystem, people and organizations are more susceptible to sophisticated assaults.

Thus, an all-in-one browser plugin that uses machine learning to offer effective spam filtering, continuous network log analysis, and real-time phishing URL identification is desperately needed. To guarantee better protection, increased productivity, and stronger network security, this solution needs to change constantly in tandem with new threats.

By resolving these issues, the suggested browser extension will contribute significantly to cybersecurity by providing a thorough and effective tool for protecting users against complex and dynamic cyberthreats.

## II. RELATED WORK

In an increasingly digital world, cybersecurity has become paramount as individuals and organizations face growing threats from malicious URLs, spam emails, and network vulnerabilities. Several notable projects and platforms have contributed to the evolution of safety measures, each addressing various types of problems through innovative approaches.

1) **Signature based Malicious URL Detection:** Studies on malicious URL detection using the signature sets had been evaluated and used long time ago [3,4,5] . Most of these studies often use lists of known malicious URLs and creates a database so whenever a new URL is encountered, a database query is executed. If the URL is blacklisted, it is considered as malicious, and then the necessary protocol is followed otherwise URLs will be considered as safe to use. The main disadvantage of this approach is that it will be very difficult to detect new malicious URLs that are not in the given database.
2) **UnMask Parasites:** Unmask Parasites [6] is a URL testing tool that analyzes URLs by downloading provided links and parsing through Hypertext Markup Language (HTML) codes, especially external links, iframes, and JavaScript. This tool's advantage is that it can detect iframes quickly and accurately. However, this tool is only useful if the user suspects something strange is happening on their sites.
3) **Dr.Web Anti-Virus Link Checker :** Dr.Web Anti-Virus Link Checker [7] is an add-on for Chrome, Firefox, Opera, and other Chromium based browsers to automatically find and scan malicious content on a download link on all social networking sites.
4) **Comodo Site Inspector :** This is a malware and security hole detection tool. This helps users check URLs or enables

webmasters to set up daily checks by downloading all the specified sites. and run them in a sandbox browser environment [8].

5) **Web of Trust (WOT):** The WOT Chrome extension [9] is a security tool that assists users in making informed choices while browsing the web. By offering a rating system for websites, it helps users in determining the trustworthiness, safety, or potential risks associated with a site. WOT's primary function is based on a reputation system that relies on input from millions of users, who contribute to website safety ratings based on their own experiences. This real-time feedback mechanism, combined with information from reliable third-party sources, adds an extra layer of protection for users against online threats like phishing, malware, and scams.

6) **Norton SafeWeb:** Norton LifeLock developed the Norton Safe Web Chrome extension [10] to provide real-time protection from harmful websites. By scanning websites for malware, phishing attempts, and other online threats, the extension alerts users to potential dangers before they access risky pages. It functions by examining website URLs and issuing safety ratings based on data collected from Norton's extensive security network. Users can see these ratings displayed as green (safe), yellow (caution), or red (dangerous) indicators in search results and when visiting websites, allowing them to steer clear of risky sites. Our Extension builds upon the strengths and addresses the limitations of these existing platforms by integrating machine-learning capabilities and a user-centric design.

## III. METHODOLOGY

The methodology for our Extension encompasses a structured approach to designing, developing, and implementing machine learning algorithms. This section outlines the key phases of the project, including requirement gathering, design and architecture, development, testing and quality assurance, and deployment and maintenance.

1) **Requirements Gathering and Analysis:** The methodology for our Extension began with an in-depth requirements gathering and analysis phase. This involved analyzing and evaluating different available security tools, to understand their working and limitations. Different datasets were considered to obtain qualitative and quantitative data on malicious URLs and their attributes.

2) **System Design and Architecture:** The next step involved designing the system architecture using a custom stack with HTML , CSS , JavaScript for the frontend, and FastAPI [11] for the back end. This architecture was selected for its reliability and efficiency in managing both front-end and back-end operations. Comprehensive design documentation was developed to outline the data flow, key system components, and integration points, ensuring a well-organized and cohesive system structure.

3) **Data Collection and Preprocessing:** To ensure the high efficiency of machine learning models we have selected the dataset [12] created by Pierre Rochet which follows all the guidelines proposed by Hannousse [13] for Phishing URL Detection. For Spam Classification spam.csv dataset [14], a widely used resource for training Machine Learning models in spam detection was employed to train our model. For Datapreprocessing methods like Steeming and Vectorization were employed.

4) **Quantitative Data: Performance and Security Evaluation:** Quantitative data plays a crucial role in evaluating the performance and reliability of the Chrome extension developed for detecting fake and malicious URLs, analyzing network logs, and classifying spam.

## IV. IMPLEMENTATION AND DEPLOYMENT

The implementation and deployment phases are crucial in bringing the Chrome extension for fake URL detection, network log analysis, and spam classification to life. These phases ensured that the system was robust, efficient, and ready for real-world use.

1) **Data Preprocessing for Model Training :** To effectively train the model, the data must be organized in a suitable format, allowing the algorithms to efficiently recognize the underlying patterns within the dataset.
For Phishing URL Detection, the chosen dataset [12] already contains the necessary features and attributes for the model. Therefore, there is no need to explicitly extract the URL's features. Instead, we simply vectorize the URL and its features before passing them to the model for training. For Spam Classification, we preprocess the data (sentences) by first converting them to lowercase and then applying tokenization to extract natural language processing (NLP) features such as the number of characters, words, and sentences. After tokenization, the words are stemmed using Porter's Stemmer algorithm [15]. The stemmed words are then rejoined and vectorized before being passed to the model for training.

2) **Selection of Classifier Algorithms :** A key aspect of utilizing Machine Learning technology is selecting an efficient classifier algorithm for the model. To achieve this, both datasets[12,14] were split in an 80:20 ratio, with 80% used for training and 20% for testing the classifier.
Various classifier algorithms, such as LinearSVC, Logistic Regression, and Random Forest, were employed for Phishing URL Detection, and their performance after training is displayed in Fig. 1.

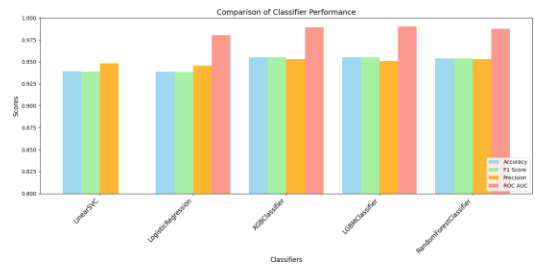

Fig. 1. Training Evaluation Metrics for Phishing URL Detection

As illustrated in Fig. 1, LGBMClassifier demonstrated the best performance, making it the chosen classifier algorithm for the Phishing URL Detection model.

For Spam Classification, a range of algorithms, including Logistic Regression, Multinomial Naive Bayes, Random Forest, and several others, were employed, and their performance after training is displayed in Fig. 2.

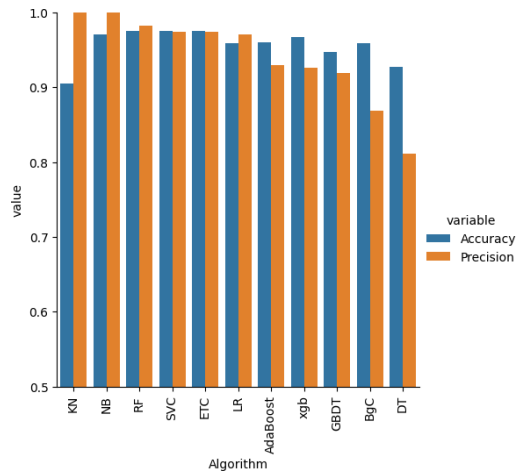

Fig. 2. Training Evaluation Metrics for Spam Classification

As illustrated in Fig. 2, Multinomial Naive Bayes demonstrated the best performance, making it the chosen classifier algorithm for the Spam Classification model.

3) **Data Preprocessing for Model Execution:** The models have been trained, but for execution in the browser extension, it is essential that the input is provided in the correct format to enable the models to function effectively. Therefore, the inputs must be structured accordingly.

In the case of the Phishing URL detection model, the URL should be accompanied by its corresponding features and attributes used during training. Additionally, this must be accomplished in a manner that does not compromise browser performance while using the model or the extension.

To achieve this, we utilized Python libraries such as Beautiful Soup [16], Googlesearch [17], and whoIs [18] to retrieve the necessary attributes.

For Spam Classification, the input message or email must be vectorized before it can be processed by the model for evaluation.

4) **Network Logs Capturing and Analysis:** To capture network logs, we utilize the webRequest API, a Chromium-specific API that enables monitoring of network activity, including HTTP requests and responses. This API allows us to block, redirect, or modify requests before they reach the web page, providing powerful tools for developing network traffic analyzers.

The HTTP status codes from the captured logs are compared against established categories of status codes to assess their threat levels. Table 1 presents the HTTP status codes encountered by Googlebot and how Google classifies each category of status code [19].

TABLE I
HTTP STATUS CODE CATEGORIZATION

| HTTP Status Codes | Status Code's Outcome |
|---|---|
| 2xx | Success |
| 3xx | Redirection |
| 4xx | Client Error |
| 5xx | Server Error |

We have created a function that runs every 30 seconds in the Browser extension's background to analyze the captured Network logs.

5) **Deployment :** The models should be deployed on the server side rather than the client side to ensure that the model's execution does not negatively impact the performance of the extension or the browser.

FastAPI was used to develop the backend for the extension, providing endpoints for both POST and GET methods. These methods facilitated communication between the extension and the models, ensuring seamless interaction.

V. EVALUATION AND RESULTS

The evaluation of the Chrome extension for fake and malicious URL detection, network log analysis, and spam classification was conducted to assess its effectiveness in meeting the defined objectives. This phase involved various testing methodologies and performance metrics to ensure that the extension functions as intended and provides a valuable user experience. The key offerings and benefits provided by the extension are outlined below:

1) **Fake and Malicious URL Detection:** The extension leverages machine learning models trained on large datasets to accurately identify fake and malicious URLs. The model analyzes domain patterns, URL structures, and other features to assess whether a website is trustworthy. This is achieved by using a LGBMClassifier algorithm which has shown the best results with the training data.

Table 2 presents the metrics of the Machine Learning model used for Phishing URL detection.

TABLE II
METRICS FOR PHISHING URL DETECTION ML MODEL

| Metrics | Score |
|---|---|
| Accuracy | 0.9650 |
| Precision | 0.9680 |
| F1 Score | 0.9680 |
| ROC AUC | 0.9931 |

The model [20] performed better than the model originally trained on the dataset [12] by more than 1.5% in each Metric.

2) **Network Log Analysis for Anomalies:** Using HTTP Status code categorization, the extension monitors network logs to detect unusual or suspicious activity in real time. This anomaly detection helps users identify potential security breaches, malware infections, or unauthorized access, contributing to a safer browsing experience. As the category of HTTP Status Code is already known there is no error in categorizing the Network Log associated with the Status Code.
3) **Spam Classification Using NLP:** The spam classifier, integrated into the extension, utilizes Natural Language Processing (NLP) techniques to analyze textual content for spam detection. The Machine Learning model for Spam Classification utilizes the Multinomial Naive Bayes Classifier which has shown the best results with the training data.

Table 3 presents the metrics of the Machine Learning model used for Spam Detection.

TABLE III
METRICS FOR SPAM DETECTION ML MODEL

| Metrics | Score |
|---|---|
| Accuracy | 0.970986 |
| Precision | 1 |

## VI. FUTURE WORK

1) **Adding a Sentiment Analyzer:** Adding a sentiment analyzer as a future feature in the extension which would enhance its functionality by analyzing the sentiment or emotional tone of textual content whixh will make it resilient to sarcastic and other out-of-context content.
2) **Adding an select to check feature:** Browser Extension for Fake URL Detection could be simplified to use by adding an select to check feature which would enable the user to use the spam classifier feature on the selected text by a mere click rather than having to paste it in the text-area.

## VII. CONCLUSION

The Chrome extension for fake and malicious URL detection, network log analysis, and spam classification represents a significant advancement in the field of cybersecurity by providing a comprehensive, user-centric tool that addresses many of the limitations found in existing online safety solutions. By integrating AI and machine learning, it offers personalized security experiences that adapt to the individual needs of users, thereby enhancing their safety and confidence while browsing the web.The extension's utilization of advanced data analytics and real-time monitoring empowers users to gain valuable insights into their online activities, fostering an environment of continuous improvement in digital safety practices. The implementation of anomaly detection in network logs further enhances its effectiveness by allowing users to identify and respond to potential threats promptly.

Moreover, the extension's commitment to flexibility and scalability ensures that it can cater to a diverse user base, accommodating various levels of cybersecurity awareness and adapting to emerging threats. Its modular design allows for ongoing enhancements and integration with cutting-edge technologies, ensuring its relevance and effectiveness in the rapidly evolving cybersecurity landscape.

REFERENCES

1) Anti-Phishing Working Group (2020). Phishing activity trends report. https://docs.apwg.org/reports/apwg_trends_report_q2_2020.pdf.
2) Wandera (2020). Mobile Threat Landscape 2020: Understanding the key trends in mobile enterprise security in 2020. Technical Report. https://www.wandera.com/mobile-threat-landscape/.
3) S. Sheng, B. Wardman, G. Warner, L. F. Cranor, J. Hong, and C. Zhang, "An empirical analysis of phishing blacklists," in Proceedings of Sixth Conference on Email and Anti-Spam (CEAS), 2009.
4) C. Seifert, I. Welch, and P. Komisarczuk, "Identification of malicious web pages with static heuristics," in Telecommunication Networks and Applications Conference, 2008. ATNAC 2008. Australasian. IEEE, 2008, pp. 91–96.
5) S. Sinha, M. Bailey, and F. Jahanian, "Shades of grey: On the effectiveness of reputation-based "blacklists"," in Malicious and Unwanted Software, 2008. MALWARE 2008. 3rd International Conference on. IEEE, 2008, pp. 57–64.
6) "UnMask Parasites". Accessed: Aug 28, 2024 [Online]. Available: https://unmask.sucuri.net/
7) "Dr.Web Anti-Virus Link Checker". Accessed: Aug 28, 2024 [Online]. Available:https://www.drweb.com/
8) "Comodo Site Inspector". Accessed: Aug 28, 2024 [Online]. Available:https://help.comodo.com/topic-208-1-490-5111-.html
9) "Web of Trust". Accessed: Aug 28, 2024 [Online]. Available: https://www.mywot.com
10) "Norton SafeWeb". Accessed: Aug 28, 2024 [Online]. Available: https://www.safeweb.norton.com
11) "FastAPI, Inc. (2024) FastAPI Documentation". Accessed: Aug 28, 2024 [Online]. Available:https://fastapi.tiangolo.com/
12) "Phishing-url dataset by Pierre Rochet". Accessed: Aug 28, 2024 [Online]. Available: https://huggingface.co/datasets/pirocheto/phishing-url
13) Hannousse & Yahiouche "Towards Benchmark Datasets for Machine Learning Based Website Phishing Detection: An Experimental Study."2020.
14) "Spam.csv dataset". Accessed: Aug 28, 2024 [Online]. Available:https://drive.google.com/file/d/1G3qVY2SbLp6rFub0Usd
15) "Porter's Steemer Algorithm". Accessed: Aug 28, 2024 [Online]. Available:https://pythonprogramming.net/stemming-nltk-tutorial
16) "BeautifulSoup". Accessed: Aug 28, 2024 [Online]. Available:https://pypi.org/project/beautifulsoup4
17) "Googlesearch". Accessed: Aug 28, 2024 [Online]. Available:https://pypi.org/project/googlesearch-python
18) "whoIs". Accessed: Aug 28, 2024 [Online]. Available:https://pypi.org/project/python-whois

19) "HTTP Status Codes". Accessed: Aug 28, 2024 [Online]. Available:https://developers.google.com/search/docs/crawling-indexing/http-network-errors
20) "ML Model". Accessed: Aug 28, 2024 [Online]. Available:https://huggingface.co/pirocheto/phishing-url-detection